\documentclass[iopscience,apj]{emulateapj}

\usepackage{placeins}
\usepackage{natbib}
\usepackage{amsmath}

\begin{document}

\newcommand\msun{M_{\odot}}
\newcommand\lsun{L_{\odot}}
\newcommand\msunyr{M_{\odot}\,{\rm yr}^{-1}}
\newcommand\be{\begin{equation}}
\newcommand\en{\end{equation}}
\newcommand\cm{\rm cm}
\newcommand\kms{\rm{\, km \, s^{-1}}}
\newcommand\K{\rm K}
\newcommand\etal{{et al}.\ }
\newcommand\sd{\partial}
\newcommand\mdot{\dot{M}}
\newcommand\rsun{R_{\odot}}
\newcommand\yr{\rm yr}

\shorttitle{ROSSBY WAVE INSTABILITY BY PROTOSTELLAR INFALL} 
\shortauthors{Bae et al.}

\title{ARE PROTOPLANETARY DISKS BORN WITH VORTICES? -- ROSSBY WAVE INSTABILITY DRIVEN BY PROTOSTELLAR INFALL}

\author{Jaehan Bae\altaffilmark{1},
Lee Hartmann\altaffilmark{1},
Zhaohuan Zhu\altaffilmark{2,3}}

\altaffiltext{1}{Deptartment of Astronomy, University of Michigan, 1085 S. University Ave.,
Ann Arbor, MI 48109, USA} 
\altaffiltext{2}{Department of Astrophysical Sciences, Princeton University,
4 Ivy Lane, Peyton Hall, Princeton, NJ 08544, USA}
\altaffiltext{3}{Hubble Fellow.}

\email{jaehbae@umich.edu, lhartm@umich.edu, zhuzh@astro.princeton.edu}

\begin{abstract}

We carry out two-fluid, two-dimensional global hydrodynamic simulations to test whether protostellar infall can trigger Rossby wave instability (RWI) in protoplanetry disks. 
Our results show that infall can trigger the RWI and generate vortices near the outer edge of the mass landing on the disk (i.e. centrifugal radius). 
We find that the RWI is triggered under a variety of conditions, although the details depend on the disk parameters and the infall pattern.
The common key feature of triggering the RWI is the steep radial gradient of the azimuthal velocity induced by the local increase in density at the outer edge of the infall region.
Vortices form when the instability enters the nonlinear regime.
In our standard model where self-gravity is neglected, vortices merge together to a single vortex within $\sim 20$ local orbital times, and the merged vortex survives for the remaining duration of the calculation ($> 170$ local orbital times).
The vortex takes part in outward angular momentum transport, with a Reynolds stress of $\lesssim10^{-2}$.
Our two-fluid calculations show that vortices efficiently trap dust particles with
stopping times of the order of the orbital time, locally enhancing the dust to gas ratio for particles of the appropriate size by a factor of $\sim 40$ in our standard model.
When self-gravity is considered, however, vortices tend to be impeded from merging and may eventually dissipate.
We conclude it may well have that protoplanetary disks have favorable conditions for vortex formation during the protostellar infall phase, which might enhance early planetary core formation.

\end{abstract}


\section{INTRODUCTION}

Dust particles in protoplanetary disks feel drag forces from the gas.
In general, the disk gas has outward pressure support and so rotates at sub-Keplerian speeds, whereas 
grains attempt to move in Keplerian motion. 
This effect is most significant for the particles that have the stopping time comparable to the orbital time, $t_s \Omega \sim 1$, where $t_s$ and $\Omega$ denotes the stopping time and the disk rotation frequency, respectively \citep{weidenschilling77}.
Radial drift of dust is thus dependent on the disk conditions under which the particles reside in, but for centimeter to meter-sized particles the resulting radial drift velocity usually reaches a few tens to hundreds m~s$^{-1}$ \citep{weidenschilling77,weidenschilling93,klahr06}.
The corresponding radial drift time at 10~AU, for instance, is less than a few $\times 10^3$ years which is much shorter than planet formation timescale as well as disk lifetime.

Formation of vortices in protoplanetary disks could be important in the context of planet formation due to their ability to efficiently trap dust particles \citep{adams95,barge95,tanga96,bracco99,chavanis00,fromang05,inaba06,heng10,birnstiel13,zhustone14,zhu14}.
Anticyclonic vortices -- with highest pressure at their centers -- are especially important since they can survive long (hundreds of orbits), while cyclonic vortices dissipate quickly \citep{godon99}.
Also, at the core of anticyclones, the radial and azimuthal gas pressure gradients vanish and therefore gas there is in Keplerian motion \citep{klahr06}.
This is an interesting feature of anticyclones because they not only concentrate dust, but once  particles reach to the vortex center there is no drag force if turbulence and scattering of particles is absent.
Therefore, vortices can be regarded as promising sites of planet formation.
In fact, highly non-axisymmetric, vortex-like structures in protoplanetary disks have been observed in submillimeter and millimeter interferometric observations: LkH$\alpha$~330, SR 21N, HD~135344B \citep{brown09}, Oph IRS~48 \citep{vandermarel13}, HD~142527 \citep{casassus13,fukagawa13}, SAO~206462, SR~21 \citep{perez14}, and PDS~70 \citep{hashimoto15}.

Possible ways to form vortices in protoplanetary disks are through the Rossby wave instability \citep[RWI;][]{lovelace99,li00,li01} or the Papaloizou-Pringle instability \citep{papaloizou84,papaloizou85,hawley87}. 
Generally, the instabilities are triggered by the additional shear arisen from the change of azimuthal velocity gradient over the radius of interest.
In two dimensions, it has been both analytically and numerically shown that the RWI can be triggered at a minimum of the generalized vortensity $\eta \equiv {\kappa^2/( 2\Sigma \Omega S^{2/\gamma})}$ \citep{lovelace99,li00}, where $\kappa$ is the epicyclic frequency, $\Sigma$ is the surface density, $\Omega$ is the rotation frequency, and $S$ is the entropy with the adiabatic index $\gamma$.
Recent numerical simulations show that the RWI appears to trigger in a similar way in three dimensions, as in two dimensions \citep{meheut10,meheut12a,lyra12,richard13,lin14}. 
The question is then how to make a vortensity minimum in protoplanetary disks?
Previously, it has been suggested that a vortensity minimum can form and the RWI is triggered at the edge of the disk dead-zone \citep{inaba06,varniere06,lyra08,lyra09,lyra12,regaly12} or at the edge of gaps carved by a planet \citep{devalborro07,lin14}.

In this paper, we propose another mechanism that possibly drives the RWI in protoplanetary disks: protostellar infall.
This is partly motivated by the recent ALMA image of HL Tau\footnote{http://www.eso.org/public/news/eso1436} showing multiple ring and gap structures which might be generated by planets.
We focus on the fact that HL Tau is embedded in an envelope from which it still accretes material \citep{beckwith89,hayashi93}. 
If the gaps in the ALMA image are indeed a signature of planets, this implies that planet formation can happen in the very early phase of protostellar evolution.
Our simulations show that protostellar infall from natal cloud can generate local vortensity minimum near the centrifugal radius, inside of which radius infalling material falls onto the disk, and can trigger the RWI. 
While the RWI activity depends on the characteristics of infall model (e.g. radial infall profile, existence of shear between infall and disk material) as well as disk parameters (e.g. viscosity parameter), we find that the RWI can be triggered under a broad circumstance and is a possible way to form vortices.
Our results suggest that vortices can form during very early evolution of protostellar systems, and this may ease the timescale problem for giant planet formation.

\section{NUMERICAL METHODS}
\label{sec:methods}

\subsection{Infall Model}
\label{sec:infall}

The infall model is implemented by adding the corresponding terms to the hydrodynamic equations as below.
\be\label{eqn:mass}
{\partial \Sigma_g \over \partial t} + \nabla \cdot (\Sigma_g v_g) = \dot{\Sigma}_{\rm in}
\en
\be\label{eqn:momentum}
\Sigma_g \left( {\partial v_g \over \partial t} + v_g \cdot \nabla v_g \right) = - \nabla P_g - \Sigma_g \nabla \Phi + \nabla \cdot \Pi_g + F_{\rm in}
\en
In the above equations $\Sigma_g$ is the gas surface density, $v_g$ is the gas velocity, $P_g=\Sigma_g c_s^2$ is the vertically integrated gas pressure, $\Phi$ is the gravitational potential including the disk self-gravitational potential (if considered), $\Pi_g$ is the viscous stress tensor for gas, respectively.
The terms $\dot{\Sigma}_{\rm in}$ and $F_{\rm in}$ indicate the changes in the equations due to the infall model, where $\dot{\Sigma}_{\rm in}$ is the mass infall rate and $F_{\rm in}$ is the shear force.
We note that in our standard model we look at the effect of density enhancement only by matching velocities of infalling material and disk material ($F_{\rm in} =0$).
The shear term $F_{\rm in}$ in the momentum equation is added later in the model where the shear in between infalling material and disk material is considered (see below and Table \ref{tab:parameter}).

In the original work of \citet{ulrich76} and \citet{cassen81}, the collapse of an isothermal, spherically symmetric, and uniformly rotating cloud was studied. 
The infalling material follows parabolic orbits, arriving at the disk surface with different radial and azimuthal velocities from those of the disk material.
Therefore, shear force exists which can be written as $F_{R,{\rm in}}=\dot{\Sigma}_{\rm in}(v_{R,\rm in}-v_{R,\rm disk})$ and $F_{\phi,{\rm in}}=\dot{\Sigma}_{\rm in}(v_{\phi,\rm in}-v_{\phi,\rm disk})$.
Here, $v_{R, \rm in}= -(GM_* / R)^{-1/2}$ and $v_{\phi,\rm in} = (GM_* / R_c)^{-1/2}$ are the velocities of the infalling material with $R_c$ being the centrifugal radius \citep{cassen81}, and $v_{R, \rm disk}$ and $v_{\phi,\rm disk}$ are the velocities of the disk, respectively.

We consider two modifications of the infall model introduced in \citet{ulrich76} and \citet{cassen81}.
First, we simplify the model in a such way that the radial and azimuthal velocities of the infalling material match those of disk material and thus no shear force exists (hereafter UCM model). 
One notable feature of the infall model of \citet{ulrich76} and \citet{cassen81} is that because of the solid-body rotation the infalling material near the rotational axis has less angular momentum and thus falls at small radius while the infalling material far from the rotational axis have more angular momentum and falls at large radius.
This will concentrate infalling material near the outer edge of the infall (i.e. centrifugal radius). 
In order to alleviate the relatively strong density enhancement around the centrifugal radius, we consider another modification (hereafter MUCM model).
In the MUCM model, the radial infall pattern is modified in a such way that mass flux per unit distance is constant over radius in order to avoid the singularity in density of the infalling material at the centrifugal radius (see below and Figure 1 of \citealt{bae13a} for comparison between UCM and MUCM model).

The mass infall rate of the UCM model is 
\be
\dot{\Sigma}_{\rm in}(R) = {{\dot{M}_{\rm in} \over {4\pi R_c  R}}}~{\left(1- {R \over R_c}\right)^{-1/2}}~{\rm if}~R \le R_c
\en
and
\be
\dot{\Sigma}_{\rm in}(R) = 0~{\rm if}~R> R_c,
\en
where $\dot{M}_{\rm in}=0.975c_{sc}^3 /G$ is the constant total infall mass rate at a given cloud isothermal sound speed $c_{sc}$ for the singular sphere solution \citep{shu77}.

In the MUCM model, the mass infall rate is smoothed as
\be
\dot{\Sigma}_{\rm in}(R) = {\dot{M}_{\rm in} \over {2\pi R_c R}}~{\rm if}~R \le R_c
\en
and
\be
\dot{\Sigma}_{\rm in}(R) = 0~{\rm if}~R> R_c.
\en

\begin{deluxetable*}{lccccccccccc}
\tablecolumns{15}
\tabletypesize{\tiny}
\tablecaption{Model Parameters \label{tab:parameter}}
\tablewidth{0pt}
\tablehead{
\colhead{Models} & 
\colhead{Case Name} & 
\colhead{$\alpha$} & 
\colhead{$R_c$} & 
\colhead{Shear Terms} & 
\colhead{Numerical Resolution} &
\colhead{Infall Model} &
\colhead{Self-gravity} \\
\colhead{} & 
\colhead{} & 
\colhead{} & 
\colhead{(AU)} & 
\colhead{} &
\colhead{($N_r \times N_\phi$)} & 
\colhead{} &
\colhead{} 
 }
\startdata
Standard Model (\S\ref{sec:standard}) & S & $10^{-4}$ & 25, fixed & N & $512\times1024$ & UCM & N \\
\hline
Numerical Resolution (\S\ref{sec:resolution}) & NR256 & $10^{-4}$ & 25, fixed & N & $256\times512$ & UCM & N \\
& NR1024 & $10^{-4}$ & 25, fixed & N & $1024\times2048$ & UCM & N \\
& NR2048 & $10^{-4}$ & 25, fixed & N & $2048\times4096$ & UCM & N \\
\hline
Viscosity (\S\ref{sec:viscosity}) & V2 & $10^{-2}$ & 25, fixed & N & $512\times1024$ & UCM & N \\
& V3 & $10^{-3}$ & 25, fixed & N & $512\times1024$ & UCM & N \\
& V5 & $10^{-5}$ & 25, fixed & N & $512\times1024$ & UCM & N \\
\hline
Increasing $R_c$ (\S\ref{sec:rc}) & IRC2 & $10^{-4}$ & 25, linearly increasing & N & $512\times1024$ & UCM & N \\
& & & 2 AU per 1000 years &  & & &   \\
& IRC5 & $10^{-4}$ & 25, linearly increasing & N & $512\times1024$ & UCM & N \\
& & & 5 AU per 1000 years &  & & &   \\
\hline
Shear Terms (\S\ref{sec:shear}) & SH & $10^{-4}$ & 25, fixed & Y & $512\times1024$ & UCM & N \\
\hline
Infall Model (\S\ref{sec:mcm}) & MUCM & $10^{-4}$ & 25, fixed & N & $512\times1024$ & MUCM & N \\
\hline
Self-gravity (\S\ref{sec:sg}) & SG & $10^{-4}$ & 25, fixed & N & $512\times1024$ & UCM & Y
\enddata
\end{deluxetable*}

\subsection{Dust Component}

In order to investigate the dust response to RWI-generated gas structures, we use the FARGO code \citep{masset00} that is modified to deal with two fluids as introduced in \citet{zhu12}.
We treat the dust component as an inviscid, pressureless fluid and dust simply feels the drag force in addition to the central stellar potential.
Note that the dust component in our calculations evolves passively and does not affect gas evolution and therefore the RWI.
Dust feedback may become important at least locally in vortices \citep[e.g.][]{fu14} but the effect is not considered in our study.

The drag terms are added in an additional source step as 
\be\label{eqn:dragr}
{\partial v_{R,d} \over \partial t} = - {v_{R,d} - v_{R,g} \over t_s}
\en
and 
\be\label{eqn:dragp}
{\partial v_{\phi,d} \over \partial t} = - {v_{\phi,d} - v_{\phi,g} \over t_s},
\en
where $v_d$ and $v_g$ denote dust and gas velocities, and $t_s$ is the dust stopping time.
With the initial setup explained below in \S\ref{sec:initial}, the mean free path of gas molecules is $\lambda = 3.8~(R/1~{\rm AU})^{4/9} \exp (R/R_c)~{\rm cm}$ so in our simulation domain the dust particles smaller than $\sim1$~m are in the Epstein regime \citep{whipple72,weidenschilling77}.
Therefore, the dust stopping time can be written as 
\be\label{eqn:tstop}
{t_s = {\rho_p s \over \rho_g v_T}},
\en
where $\rho_p$ is the dust particle density, $s$ is the dust particle radius, $\rho_g$ is the gas density, and $v_T=\sqrt{8/\pi} c_s$ is the mean thermal velocity with $c_s$ being the gas sound speed \citep{takeuchi02}.
The dust particle density is assumed to be $\rho_p=1~\rm{g~cm}^{-3}$ in this study.
With the two-dimensional approach the gas mass density is $\rho_g = \Sigma_g / \sqrt{2\pi} H$ and thus the dust stopping time can also be written as 
\be\label{eqn:tstop2}
{t_s = {\pi \rho_p s \over 2 \Sigma_g \Omega}},
\en
where $\Omega$ is the Keplerian angular velocity.
The stokes number, or the nondimensional stopping time, $T_s$ is then
\be\label{eqn:Tstop}
{T_s = t_s \Omega}.
\en

In this work, we consider two different dust sizes: small dust particles that have $T_s \sim 0.1$ and large dust particles that have $T_s \sim 1$, under the initial conditions explained in the next section.
In physical size, the former and latter corresponds to 1~cm and 10~cm, respectively.

\subsection{Initial Conditions}
\label{sec:initial}

We begin with a $0.5~\msun$ central protostar and a surrounding disk having an initial gas density distribution of $\Sigma(R) = 1000~(R/{\rm AU})^{-1} {\exp(-R/R_c)} ~{\rm g~cm^{-2}}$.
Here, $R_c$ is the centrifugal radius which is fixed to 25~AU during the calculations unless otherwise stated.
We assume a constant infall rate of $3.0 \times 10^{-6}~\msunyr$ which corresponds to a singular isothermal collapse \citep{shu77} of a 16~K protostellar cloud. 
The disk mass is $0.014~\msun$ initially and is $\sim0.11-0.13~\msun$ at the end of calculations, depending on model parameters.
We adopt a fixed radial temperature distribution ($T \propto R^{-1/2}$) that corresponds to a ratio of disk scale height to radius $H/R = 0.05~(R/1~{\rm AU})^{0.25}$.
Initial random perturbation is applied to surface density at the level of $10^{-4}$.

We use inner and outer boundaries of 5~AU and 100~AU and adopt 512 logarithmically spaced radial grid-cells and 1024 linearly spaced azimuthal grid-cells.
With this choice, $\Delta R /R$ is constant to 0.006 and grid-cells have comparable radial and azimuthal size at all radii.
In \S\ref{sec:resolution}, we discuss the effect of numerical resolution and show the results converge resolutions at $512\times1024$ and beyond.

We use $\alpha=10^{-4}$ for our fiducial viscosity parameter.
Using a relatively low disk viscosity is motivated by the fact that stellar X-rays and FUV photons are likely to be blocked by infalling material during the protostellar phase so that ionization level at disk surface layers remains low, limiting mass transport through the magnetorotational instability.
The effect of the viscosity parameter is tested in \S\ref{sec:viscosity}.
Model parameters are summarized in Table \ref{tab:parameter}.

\subsection{Boundary Conditions}

In order to prevent unphysical, rapid depletion of material at the inner boundary, we implement a velocity limiter as introduced in \citet{pierens08} and \citet{zhu12}.
To briefly summarize, we limit the radial velocities of the gas component at the inner boundary to be no more than some factors of the viscous radial velocity in a steady state, $v_{R, {\rm vis}}$: $v_{R,g} \leq \beta v_{R,{\rm vis}}$.
Here, $v_{R,{\rm vis}} = - 3 \nu_{\rm in} / 2 R_{\rm in}$ where $\nu_{\rm in}$ and $R_{\rm in}$ are the viscosity and radius at the inner boundary.
Based on a set of experiments, we find that $\beta=3$ is most suitable for this study.

A similar approach is implemented for the dust component, following \citet{zhu12}.
In case of dust, we limit radial velocity to be no more than three times of the dust drift speed in a viscous disk $v_{R, {\rm drift}}$, where $v_{R, {\rm drift}}$ is defined as  
\be
v_{R,{\rm drift}} = - {(3\nu_{\rm in}/2R_{\rm in}) T_s^{-1} + \eta v_K\over T_s + T_s^{-1}}.
\en
Here, $\eta$ is the ratio between the pressure gradient and gravitational force defined as $\eta \equiv -(R \Omega^2 \rho_g )^{-1} \partial P / \partial R$, and $v_K$ is the Keplerian speed.

At the outer boundary, standard open boundary conditions are implemented for both gas and dust components.

\section{RESULTS}
\label{sec:results}

\subsection{Standard Model: Results With the UCM Model}
\label{sec:standard}

We start with our standard model, showing details of the launching, growth, and saturation of the RWI and dust's response.
Then, in the following subsections we present how varying disk conditions and infall pattern affect the RWI activity.

\begin{figure}
\centering
\epsscale{1.2}
\plotone{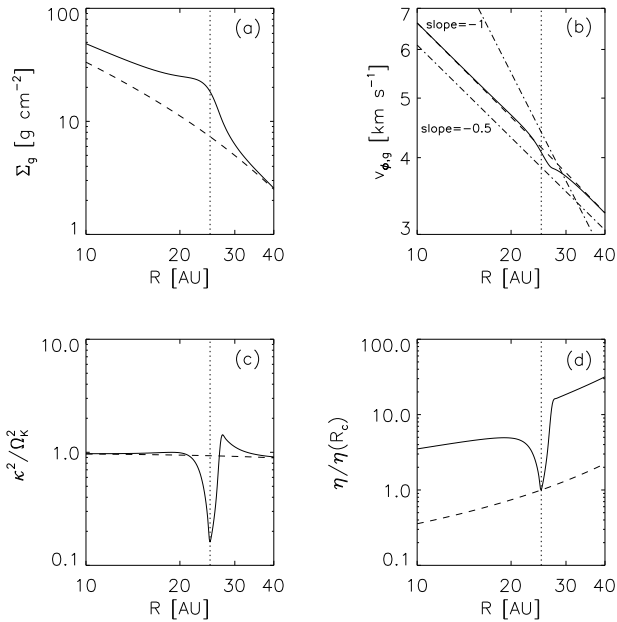}
\caption{Radial distributions of azimuthally-averaged (a) gas surface density $\Sigma_g$, (b) gas azimuthal velocity $v_{\phi,g}$, (c) epicyclic frequency $\kappa^2/\Omega_K^2$, and (d) vortensity $\eta/\eta(R_c)$ at the launching of the RWI ($t=14$~orbital times at $R_c$; see Figure \ref{fig:growth}) for the standard model. Dashed curves show the initial distributions. The centrifugal radius $R_c$ is indicated with vertical dotted lines. In panel (b), two dotted-dashed lines show slopes of $-1$ and $-0.5$, respectively. We note that the density bump generates a steep azimuthal velocity gradient around $R_c$ in order to maintain the radial force balance, which results in the $\kappa^2$ and $\eta$ minima.}
\label{fig:launching}
\end{figure}

\begin{figure}
\centering
\epsscale{1.2}
\plotone{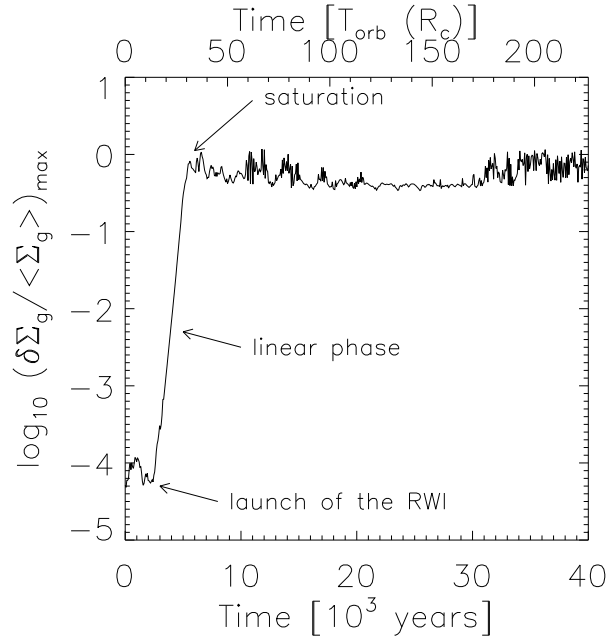}
\caption{The maximum value of non-axisymmetric gas density perturbation $\delta \Sigma_g / \langle \Sigma_g \rangle$ is shown as a function of time. The instability triggers at $14~T_{\rm orb}$ and grows exponentially during the `linear' phase ($14~T_{\rm orb} \lesssim t \lesssim 30~T_{\rm orb}$). Then the instability saturates and turns into the nonlinear regime thereafter.}
\label{fig:growth}
\end{figure}

\subsubsection{Launching of the RWI}
\label{sec:launch}

In two-dimensions, one can analytically show that the RWI can be triggered at a radial minimum of the generalized vortensity $\eta$, where
\begin{eqnarray}
\eta  = {\kappa^2 \over 2 \Sigma \Omega} {1 \over S^{2/\gamma}}
\end{eqnarray}
\citep{lovelace99,li00}. 
Here, $\kappa=[R^{-3} d(R^4 \Omega^2)/dR]^{1/2}$ is the epicyclic frequency, $\Sigma$ is the surface density, $\Omega$ is the rotational frequency, and $S=P/\Sigma^\gamma$ is the entropy.
Given that we assume a locally isothermal temperature profile which is unchanged over time, the above equation can be written as 
\begin{eqnarray}
\label{eqn:eta}
\eta  =  {\kappa^2 \over 2 \Sigma \Omega} {1 \over c_s^4}.
\end{eqnarray}
For convenience we call the quantity $\eta$ as vortensity hereafter.

Figure \ref{fig:launching} shows radial distributions of azimuthally-averaged gas surface density, azimuthal velocity, epicyclic frequency, and vortensity at the time of the launching of the RWI.
The density gradient and thus the pressure gradient becomes steeper on the outer part of the density bump as infall adds mass onto the disk.
The steep pressure gradient then generates a steeper azimuthal velocity profile in order to maintain the radial force balance.
The resulting azimuthal velocity slope deviates from $-0.5$, which is for the Keplerian rotation, and approaches to $-1$.
We note that the $-1$ slope is important in triggering the RWI since $\kappa^2$ and $\eta$ change sign at $v_\phi \propto R^{-1}$.
This can be simply shown by substituting $\kappa^2$ into Equation (\ref{eqn:eta}):
\begin{eqnarray}
\label{eqn:eta1}
\eta = {1 \over 2\Sigma \Omega} {1 \over c_s^4} {1 \over R^{3}} {d \over dR}(R^2 v_\phi^2).
\end{eqnarray}
In Keplerian disks where $v_\phi \propto R^{-1/2}$, $\eta$ is always positive.
However, $\eta$ becomes zero if $v_\phi \propto R^{-1}$ and is negative if $v_\phi$ has a steeper slope than $-1$.

\begin{figure*}
\centering
\epsscale{1.2}
\plotone{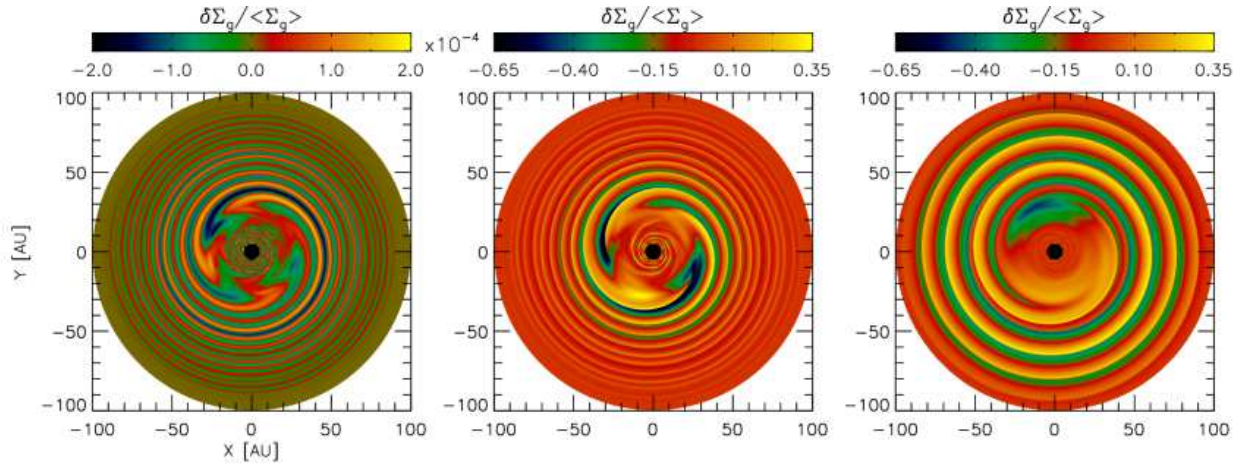}
\caption{Perturbed gas density distributions $\delta \Sigma_g/ \langle \Sigma_g \rangle$ of the standard model at the launching of the instability, at the saturation, and at the end of the simulation (from left to right). These correspond to 14, 30, and 226 local orbital times at $R_c$ or $t=2.5, 5.3$, and $40\times10^3$ years. Note that the scale for the leftmost panel is different from the other two.}
\label{fig:relden}
\end{figure*}

Figure \ref{fig:growth} presents the maximum value of non-axisymmetric density perturbation $\delta \Sigma_g / \langle \Sigma_g \rangle$ as a function of time, where $\delta \Sigma_g = \Sigma_g - \langle \Sigma_g \rangle$ and the brackets denote the azimuthal average. 
Initially before the RWI triggers, the maximum perturbed density maintains $\sim10^{-4}$ which corresponds to the initial random component.
The vortensity minimum develops gradually until the instability triggers at $t \sim 14~T_{\rm orb}$, where $T_{\rm orb}$ hereafter denotes the local orbital time at the centrifugal radius. 

\subsubsection{Growth and Saturation of the RWI}
\label{sec:growth}

After the instability is triggered, it grows in the `linear' regime where the growth of the perturbed density is well described by the linearized continuity equation $\partial (\delta \Sigma) / \partial t = - i \tilde{\omega} \delta \Sigma$, where $\tilde{\omega} = \omega + i \gamma$ is a complex frequency, $\omega$ is the real mode frequency, and $\gamma$ is the growth rate.
This is clearly seen in Figure \ref{fig:growth} -- the instability grows exponentially during the linear phase ($14~T_{\rm orb} \lesssim t \lesssim 30~T_{\rm orb}$). 
The growth timescale $T_{\rm growth}$, during which time the maximum perturbed density increases by a factor of $e$, is $1.8~T_{\rm orb}$.
After spending $16~T_{\rm orb}$ in the linear phase, the instability saturates and enters the nonlinear regime.

Figure \ref{fig:relden} shows the spatial distributions of the perturbed gas density at the launching of the instability ($t=14~T_{\rm orb}$), at its saturation ($t=30~T_{\rm orb}$), and at the end of the simulation ($t=226~T_{\rm orb}$).
Initially, the $m=4$ mode is dominant but it quickly merges to $m=3$ mode during the linear phase, and eventually merges to $m=1$ mode which is maintained until the end of the calculation.

\begin{deluxetable}{lccccccccccc}
\tablecolumns{15}
\tabletypesize{\scriptsize}
\tablecaption{Results\label{tab:result}}
\tablewidth{0pt}
\tablehead{
\colhead{Case Name} &
\colhead{RWI} &
\colhead{$\kappa^2/\Omega_K^2$\tablenotemark{a}} &
\colhead{$T_{\rm launch}$\tablenotemark{b}} &
\colhead{$T_{\rm sat}$\tablenotemark{b}} &
\colhead{$T_{\rm growth}$\tablenotemark{b}} &
\colhead{Vortex} \\
\colhead{} &
\colhead{} &
\colhead{} &
\colhead{($T_{\rm orb}$)} &
\colhead{($T_{\rm orb}$)} &
\colhead{($T_{\rm orb}$)} &
\colhead{Formation} 
 }
\startdata
S & Y  & 0.16 & 14 & 30 & 1.8 & Y\\
\hline
NR256 & Y & 0.24 & 13 & 29 & 2.1 & Y \\
NR1024 & Y & 0.15 &14 & 30 & 1.7 & Y\\
NR2048 & Y & 0.15 &16 & 32 & 1.7 & Y\\
\hline
V2 & Y\tablenotemark{c} & 0.62 & 22 & - & 2.2 & N\\
V3 & Y & 0.25 & 18 & 36 & 2.1 & Y\\
V5 & Y & 0.15 & 14 & 30 & 1.8 & Y\\
\hline
IRC2 & Y & 0.28 & 22 & 33 & 2.2 & Y\\
IRC5 & Y\tablenotemark{c} & 0.60 & 24 & - & 5.3 & N\\
\hline
SH & Y & 0.03 & 8 & 19 & 1.2 & Y\\
\hline
MUCM & Y & 0.35 & 54 & 110 & 5.5 & Y\\
\hline
SG & Y & 0.15 & 14 & 32 & 1.9 & Y\tablenotemark{d}
\enddata
\tablenotetext{a}{The $\kappa^2 / \Omega_K^2$ values are measured at the launching of the RWI.}
\tablenotetext{b}{The times correspond to the launching of the RWI ($T_{\rm launch}$), the saturation of the RWI ($T_{\rm sat}$), and the exponential growing timescale ($T_{\rm growth}$).}
\tablenotetext{c}{The instability is triggered but stays only in the linear regime and withers away before it enters the nonlinear phase.}
\tablenotetext{d}{Multiple vortices form as the RWI enters the nonlinear regime, but they later dissipate as the disk becomes gravitationally unstable.}
\end{deluxetable}

\subsubsection{Vortex Formation and Dust Response}
\label{sec:dust}

\begin{figure*}
\centering
\epsscale{1.05}
\plotone{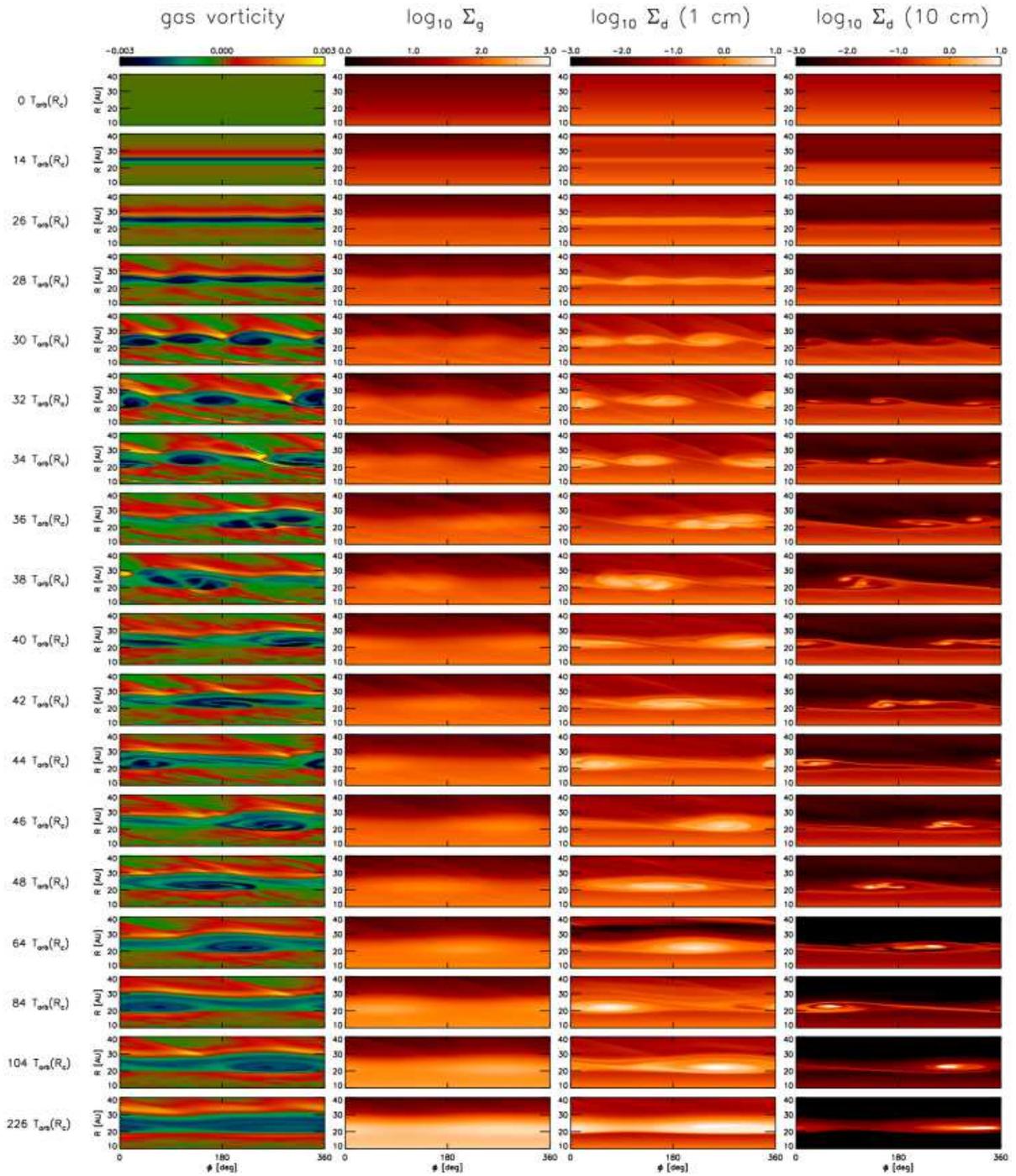}
\caption{Snapshots of gas vorticity $\nabla \times v_g$, gas density, 1~cm particle density, 10~cm particle density at selected times. Times are presented on the left side in units of $T_{\rm orb}$. The Keplerian component in the azimuthal velocity is subtracted when calculating the vorticity. Densities are in cgs units and displayed in the logarithmic scale.}
\label{fig:snap}
\end{figure*}

The RWI accompanies vortex formation as the instability enters the nonlinear regime.
Figure \ref{fig:snap} illustrates gas vorticity, gas surface density, and surface densities of 1~cm and 10~cm dust particles on the $\phi-R$ coordinates.
The (negative) vorticity minimum grows at $R_c$ as infall proceeds.
At the time the RWI initiates ($t=14~T_{\rm orb}$), the non-axisymmetric features are still too small to be seen.
As the instability grows, the radial vorticity minimum starts to develop structure and finally breaks into vortices when the instability enters the nonlinear regime ($t=30~T_{\rm orb}$).
Vortex formation during nonlinear evolution of the RWI is in good agreement with previous hydrodynamic simulations \citep[e.g.][]{li01}.
The important feature is that the vortices have a local vorticity minimum at its center -- the vortices are anticyclones.
The vortices wander in azimuth, merge together, and form a single vortex within $\sim 20~T_{\rm orb}$ from their formation.
The merged vortex survives until the end of the calculation and we did not follow its evolution thereafter.

Before the vortices form, dust particles are concentrated at the vorticity minimum while the small and the large dust behave somewhat differently.
Initially when the instability grows in the linear phase, small dust concentrate around the gas pressure maximum.
On the other hand, the large particles have the greatest inward drift velocity at the centrifugal radius ($T_s \sim 1$ at $R_c$), so the density drops at the radius.
The large dust distribution shows a steep density change around $R_c$ (see Figure \ref{fig:snap} at $t=14$ and $26~T_{\rm orb}$).
This is because infall adds gas inside $R_c$ and thus the dust stopping time correspondingly decreases significantly at the region, slowing down the large dust migrating inward.

After the vortices form, they efficiently trap dust particles so that the dust distributions show stronger asymmetry than the gas distribution.
While the small particles are well coupled to gas and follow the gas vorticity distribution well, the large particles show highly perturbed behavior.
The accumulated cores of large dust are often offset from the vortex cores.
This is more significant in earlier times when the vortices merge (see Figure \ref{fig:snap} at $t=38$ and $40~T_{\rm orb}$ for example).
Also, the large dust particles show more dramatic concentration than small dust particles.
We will discuss dust trapping and its implication later in \S\ref{sec:dust_trapping}.

\begin{figure*}
\centering
\epsscale{1.1}
\plotone{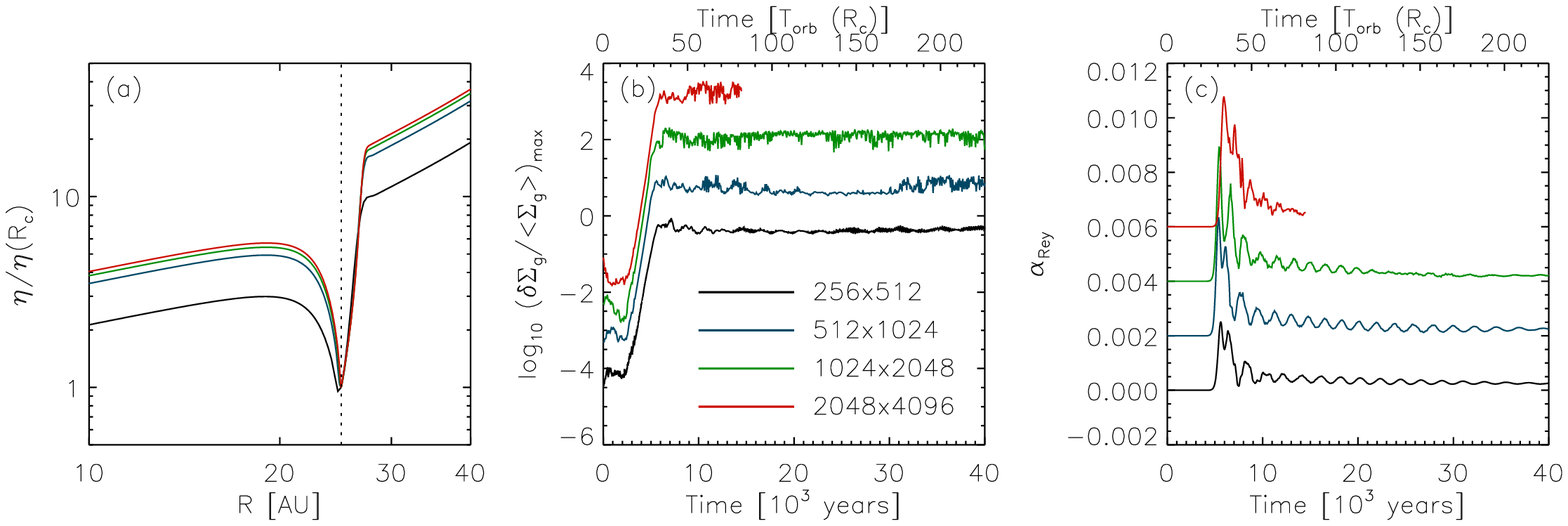}
\caption{(a) Radial distribution of the azimuthally-averaged $\eta/\eta(R_c)$ at $t=14~T_{\rm orb}$. (b) Maximum perturbed gas density $\delta \Sigma_g/ \langle \Sigma_g \rangle$ as a function of time. (c) The Reynolds stress as a function of time. In panel (b) and (c), the plots are shifted vertically by 1 and 0.002, respectively, for better view. We note that the oscillating feature of $\alpha_{\rm Rey}$ is due to the epicyclic motion of gas inside vortices against the geometric center of the vortices. Since it requires a lot of computational resources the highest resolution run is conducted only for $80~T_{\rm orb}$.}
\label{fig:res}
\end{figure*}

\begin{figure}
\centering
\epsscale{1.2}
\plotone{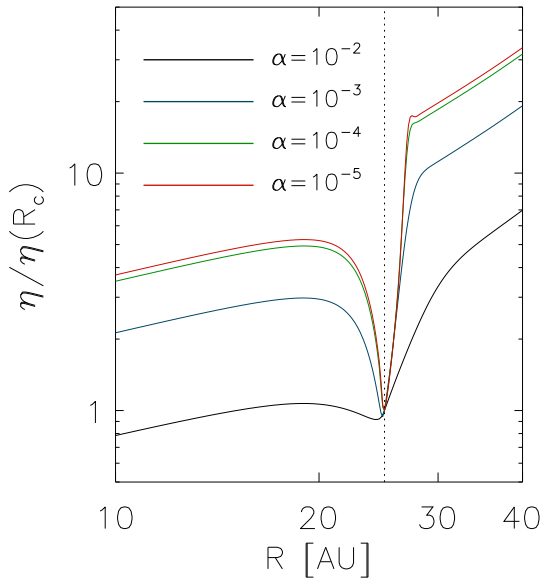}
\caption{Radial distribution of azimuthally-averaged $\eta/\eta(R_c)$ at $t=14~T_{\rm orb}$ with $\alpha=10^{-2}$, $10^{-3}$, $10^{-4}$ (standard model), and $10^{-5}$. The vortensity minimum is very shallow and broad with $\alpha=10^{-2}$. In this case, the RWI grows only in the linear regime and does not further develop to the nonlinear regime.}
\label{fig:viscosity}
\end{figure}

\begin{figure}
\centering
\epsscale{1.2}
\plotone{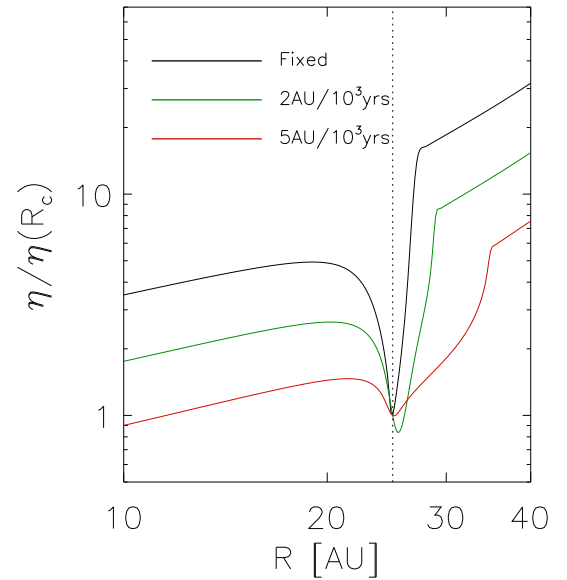}
\caption{Radial distributions of the azimuthally-averaged $\eta/\eta(R_c)$ with fixed $R_c$ at 25~AU (standard model) and linearly increasing $R_c$ at rates of 2~AU and 5~AU per 1000~years. Although the vortensity minimum is seen in all models, the RWI grows only in the linear regime when the centrifugal radius increases 5~AU per 1000~years.}
\label{fig:rc}
\end{figure}

\subsection{Effect of Numerical Resolution}
\label{sec:resolution}

As mentioned in \S\ref{sec:infall}, the UCM model has a singularity (infinite mass infall rate per unit area) at the centrifugal radius while the total infall rate is finite.
With finite grids this can cause non-convergent behavior at different numerical resolutions.
Thus, we need to show that our results are not dependent on the numerical resolution before we go to further analysis.
We test with four different numerical resolutions of $(N_R, N_\phi) = (256, 512), (512, 1024), (1024, 2048)$, and $(2048, 4096)$.

We use three different diagnostics to check the numerical convergence.
First, we check the position, shape, and depth of the vortensity minimum since it is crucial for triggering the RWI.
Second, we use the maximum value of the perturbed gas density.
For the last, we calculate the Reynolds stress which is defined as
\be
\alpha_{\rm Rey} = {{ \int {\Sigma_g \delta v_{R,g} \delta v_{\phi,g}}~dS } \over {\int {\Sigma_g c_s^2}~dS}},
\en
where the integration is done in a volume-averaging manner, and $\delta v_{R,g} = v_{R,g} - \langle v_{R,g} \rangle$ and $\delta v_{\phi,g} = v_{\phi,g} - \langle v_{\phi,g} \rangle$.
The first two quantities check the convergence of local features whereas $\alpha_{\rm Rey}$ checks the numerical convergence in a globally-averaged sense.

The three diagnostics are presented in Figure \ref{fig:res}.
All diagnostics converge well at $512 \times 1024$ and beyond.
With $256 \times 512$ grid cells, the region around the $R_c$ is not very well resolved so that the infall rate seems underestimated at the region.
The vortensity minimum with $256 \times 512$ grids is significantly shallower and broader than others (see also Table \ref{tab:result} for the minimum $\kappa^2/\Omega_K^2$ values), and thus the instability growth and the Reynolds stress are less significant.

\subsection{Effect of Viscosity}
\label{sec:viscosity}

Since larger gas viscosity more efficiently spreads out the density enhancement, it is possible to broaden the vortensity minimum and limit the RWI growth.
We tested different viscosity parameters to see the effect of disk viscosity; $\alpha = 10^{-2},10^{-3},10^{-4}$, and $10^{-5}$.
As expected, Figure \ref{fig:viscosity} shows that the vortensity minimum becomes broader as $\alpha$ increases.
Especially with $\alpha=10^{-2}$, the minimum is very shallow and broad.
In this case, the RWI triggers but it only stays in the linear regime and does not turn into the nonlinear regime -- we therefore do not observe any vortices forming and concentration of dust particles.

It has been empirically shown that the width of the vortensity minimum has to be $\lesssim 2H$ in order for the RWI to develop \citep{lyra09,regaly12}.
If we adopt this criterion, the RWI growth can be limited if the local viscous timescale at $R_c$, in which time the disk can viscously spread out the bump by $\sim H$, is shorter than the RWI growing timescale. 
The local viscous timescale estimated at $R_c$ is $t_\nu = H^2/\nu \sim (0.11R)^2/\nu \sim 90 / \alpha$~years.
With $\alpha=10^{-2}$, the local viscous timescale is $\sim 9000$~years (or $\sim 51~T_{\rm orb}$) which is approximately the timescale of the linearly growing part of the RWI.
Thus, it makes sense that the instability with $\alpha=10^{-2}$ does not further grow and turn into the nonlinear regime.
On the other hand, viscosity does not affect triggering of the RWI when the viscous timescale is 
much longer than the RWI growing timescale.

\begin{figure}
\centering
\epsscale{1.2}
\plotone{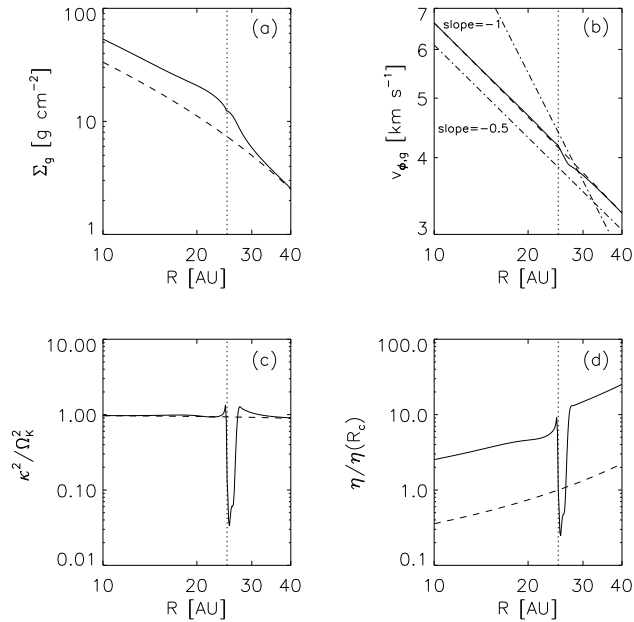}
\caption{Same as Figure \ref{fig:launching} but at $t=8~T_{\rm orb}$ for the SH model (shear terms are included). We emphasize that the vortensity minimum is much sharper and narrower compared to the standard model (see Figure \ref{fig:launching}d).}
\label{fig:vort_shear}
\end{figure}

\subsection{Effect of Linearly Increasing Centrifugal Radius With Time}
\label{sec:rc}

In protoplanetary disks, the centrifugal radius will generally not be constant with time as material with even greater angular momentum falls in; this will spread the vortensity minimum.
Instead of implementing a more self-consistent time evolution of the centrifugal radius \citep[e.g.][]{cassen81}, we mimic the effect by simply linearly increasing $R_c$ as a function of time.
We test with two different rates: $R_c$ increases at 2~AU and 5~AU per 1000~years.

In order to see how the $R_c$ increasing rates translate to initial core rotation, let us assume that the protostellar system we consider in this study evolves from the initial Bonner-Ebert sphere-like two-component density profile \citep[see][]{zhu10}, and the inner flat core has collapsed to a $0.1~\msun$ central protostar and the rest of the cloud collapses as a rotating singular isothermal sphere.
If we further assume that the disk mass inside of our inner boundary ($R_{\rm in} = 5$~AU) is negligible, our model is $\sim 1.3\times10^5$~years past from the initial core collapse.
Then, the increasing rates of $R_c$ we adopt here correspond to the instantaneous $R_c$ increasing rate of uniformly rotating cores at $\Omega_c = 5.3\times10^{-14}$ and $8.3\times10^{-14}~{\rm rad~s^{-1}}$.
The rotation frequencies are 20 and $30~\%$ of the breakup angular frequency at the outer edge of an $1~\msun$ cloud.
It is worth to point out that the rotation frequencies are large compared to the median value of $3~\%$ inferred by \citet{bae13b}; the value was required to reproduce the observed circumstellar disk frequencies as a function of age, assuming disk dispersal by photoevaporation.

Figure \ref{fig:rc} shows radial distributions of vortensity minimum for the models.
The RWI excites and generates vortices with small increasing rate of 2~AU per 1000~years.
However, if $R_c$ increases fast at 5~AU per 1000~years the RWI triggers but stays only in the linear regime.
The instability with the large rate eventually withers away and no vortices form in this case.
Although a more self-consistently evolved model is require to conclude, in very rapidly-rotating systems the centrifugal radius may move outward so fast that the RWI may not have chance to form vortices.

\subsection{Effect of Shear Terms}
\label{sec:shear}

In actual protostellar systems, the infalling material must have different specific angular momentum or azimuthal velocity than the disk material when it lands on the disk. 
Otherwise, the mass would not fall in.
We test the effect of shear by adopting velocity fields of infalling material that follows parabolic orbits.
Following \citet{ulrich76} and \citet{cassen81}, the infalling material has radial and azimuthal velocities of $v_{R, \rm in}= -(GM_* / R)^{-1/2}$ and $v_{\phi,\rm in} = (GM_* / R_c)^{-1/2}$.
So at $R=R_c$ both radial and azimuthal velocities of infalling material are the same as the Keplerian velocity at the radius.
We note that the radial velocity of infalling material is extremely large when compared to the accretion velocity in the steady-state $\alpha$ disk.
In an $\alpha$ disk, the accretion rate can be described as $\dot{M}_{\rm acc} \simeq 3 \pi \nu \Sigma$ since we are interested in the region far from the stellar surface.
Then, if we relate it to $\dot{M}_{\rm acc} = - 2\pi R \Sigma v_{R,\rm{disk}}$ we get $v_{R,\rm{disk}} = - (3/2) \alpha c_s (H/R)$.
Thus, $v_{R, \rm{disk}}/v_{R,\rm{in}} = (3/2) \alpha (H/R)^2 \ll 1$.

\begin{figure*}
\centering
\epsscale{1.2}
\plotone{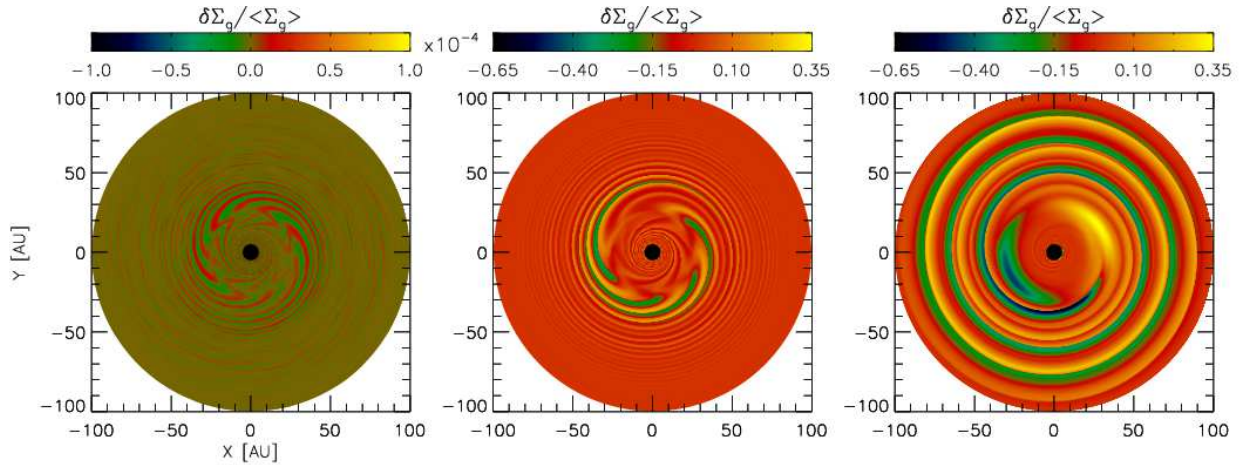}
\caption{Perturbed gas density $\delta \Sigma_g/ \langle \Sigma_g \rangle$ distributions at the launching of the instability, at the saturation, and at the end of the simulation (from left to right) for the SH model (shear terms are included). These correspond to $t = 8, 19,$ and $226~T_{\rm orb}$ or $1.4, 3.4$, and $40\times10^3$~years. Note that the scale for the leftmost panel differs from the other two.}
\label{fig:figpolarcomp_shear}
\end{figure*}

Due to the large inward radial velocity of the infalling material, the disk develops a very sharp density jump at the centrifugal radius.
Therefore, compared to the standard model where no shear terms are included, the vortensity minimum is much sharper and narrower as seen in Figure \ref{fig:vort_shear}.
In this model the shear from infall rapidly builds the vortensity minimum, whereas in the standard model the disk has to wait until enough mass is added to build the density bump.
Thus, no evident density bump is developed around $R_c$ at the time of RWI initiation.
Another notable feature is that the RWI initially triggers with an extremely high order mode of $m=9$ as seen in Figure \ref{fig:figpolarcomp_shear}, at which mode the linear growth rate is the highest in this model.
The perturbed density peaks rotate at slightly different velocities and thus one catches another as time goes.
They eventually merge to $m=1$ mode at $t\sim 100~T_{\rm orb}$.

\subsection{Effect of Infall Profile: With the Modified UCM Model}
\label{sec:mcm}

\begin{figure}
\centering
\epsscale{1.2}
\plotone{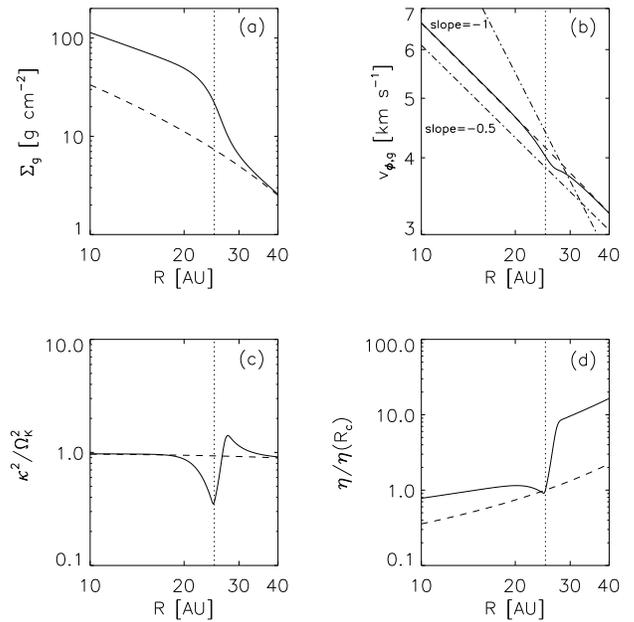}
\caption{Same as Figure \ref{fig:launching} but for the MUCM model. }
\label{fig:vort_infall}
\end{figure}

The UCM model, as pointed out earlier, adds a large fraction of infalling material near the centrifugal radius.
In order to see if the RWI can be limited by more gentle infall pattern, we match the radial infall pattern to the initial disk surface density distribution ($\Sigma_g, \dot{\Sigma}_{\rm in} \propto R^{-1}$).
Even with the smoothed infall profile we find that the RWI excites.
Figure \ref{fig:vort_infall} presents radial distributions of azimuthally-averaged gas surface density, azimuthal velocity, epicyclic frequency, and vortensity at the time of the launching of the RWI ($t=54~T_{\rm orb}$).
The vortensity minimum is shallow and broad, but since infall keeps adding material at the same radius the RWI triggers.
The density around $R_c$ increases steeply but does not develop bumpy structures.
We note that the instability very slowly grows, having three times longer $T_{\rm growth}$ compared to the standard model (see Table \ref{tab:result}).

\subsection{Effect of Self-gravity}
\label{sec:sg}

In our standard run, the azimuthally-averaged Toomre $Q$ parameter around $R_c$ is $\lesssim 1$ at $T \gtrsim 200~T_{\rm orb}$.
At this point (or even earlier), we expect disk self-gravity becomes important and may alter the later disk evolution including possible activation of gravitational instability (GI).
We thus conducted a calculation with disk self-gravity included, using FARGO-ADSG code \citep{baruteau08}.

\begin{figure*}
\centering
\epsscale{1.2}
\plotone{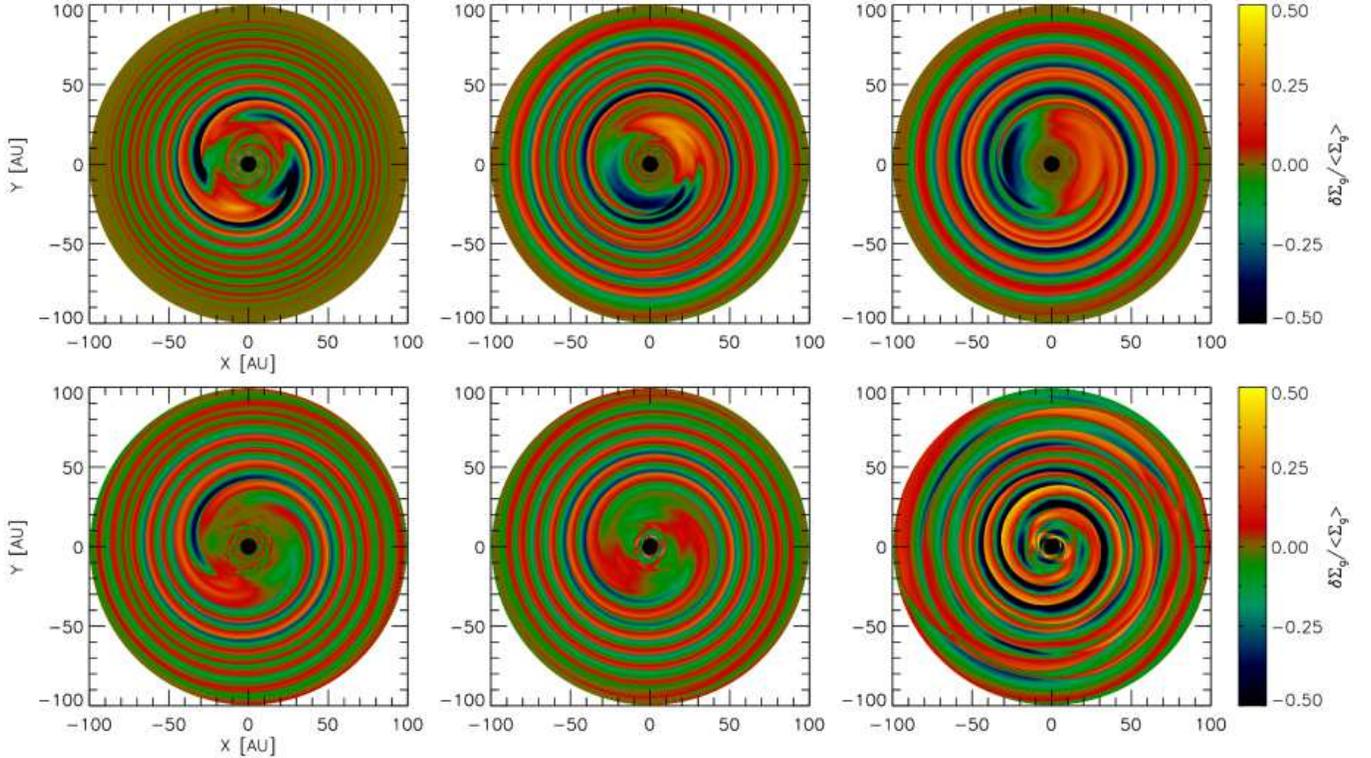}
\caption{Perturbed gas density $\delta \Sigma_g/ \langle \Sigma_g \rangle$ distributions at $t = 30, 44,$ and $160~T_{\rm orb}$ (from left to right) for the standard run (upper panels) and the self-gravity run (lower panels). The azimuthally-averaged Toomre $Q$ parameters of the self-gravity run at the three epochs are 5.3, 3.2, and 1.3.}
\label{fig:relden_sg}
\end{figure*}

Before looking at numerical results, one may predict the role of self-gravity in triggering the RWI through a simple back-of-the-envelope calculation using the connection between the vortensity $\eta$ and the Toomre $Q$ parameter, where the vortensity is again 
\be
\eta = {\kappa^2 \over 2\Omega \Sigma} {1 \over c_s^4},
\en
and the Toomre $Q$ parameter is 
\be
Q = {\kappa c_s \over \pi G \Sigma}.
\en
At a given radii, under the locally isothermal assumption, $\eta \propto \kappa^2/\Sigma$ and $Q \propto \kappa/\Sigma$.
Remember that both RWI and GI acts in a way to redistribute mass in the disk so the disk stabilizes against the instabilities. 
In other words, the instabilities broaden the density enhancement and increase the epicyclic frequency close to Keplerian speeds.
If a disk is under the circumstance that RWI and GI competes, this process will increase $\eta$ faster than $Q$ so that the RWI will stabilize first.
Therefore, one can expect RWI will still trigger while a disk is gravitationally stable ($Q \gg 1$) but as the disk becomes gravitationally unstable ($Q \sim 1$) GI can eventually take part in, in which case the RWI can be quenched.

Figure \ref{fig:relden_sg} depicts perturbed density distributions at three selected times ($t= 30,44$, and $160~T_{\rm orb}$), together with the distributions of the standard run for comparison. 
During the period the RWI linearly evolves, the azimuthally-averaged Toomre $Q$ parameter in the disk is $> 5$; the disk is gravitationally stable.
The RWI thus triggers as we predict above and vortices form.
Vortices form as the RWI enters the non-linear regime and merge together over time.
However, in the presence of self-gravity the merging tends to be impeded and results in $m=2$ mode, instead of a single vortex.
At $t=44~T_{\rm orb}$, azimuthally-averaged the Toomre $Q$ parameter is 3.2 at the radial density bump while $Q$ is locally as small as 2.4 at the core of the vortices.
As the disk further obtains material and becomes gravitationally unstable, the vortices dissipate and the GI eventually triggers generating strong $m=2$ trailing spiral arms.
The evolution of vortices under the influence of self-gravity seen in our calculation agrees well with previous two- and three-dimensional simulations showing that vortex merging can be delayed in weakly self-gravitating disks and global spiral waves develop instead of vortex formation if self-gravity is sufficiently strong \citep{lin11,lin12}.  

\begin{figure}
\centering
\epsscale{1.2}
\plotone{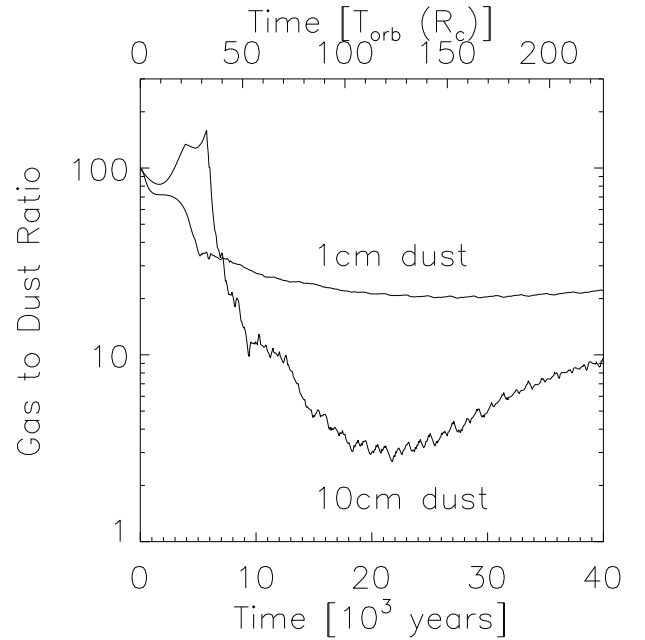}
\caption{Time evolution of the minimum gas to dust ratio for the standard model. The ratio drops significantly as the instability enters the nonlinear regime and vortices form. Because the large dust at the outer disk are depleted due to rapid radial drift and there are no more supply from the region, the ratio for the 10~cm particles after $\sim 120~T_{\rm orb}$ increases.}
\label{fig:gtd}
\end{figure}

\begin{figure*}
\centering
\epsscale{1.2}
\plotone{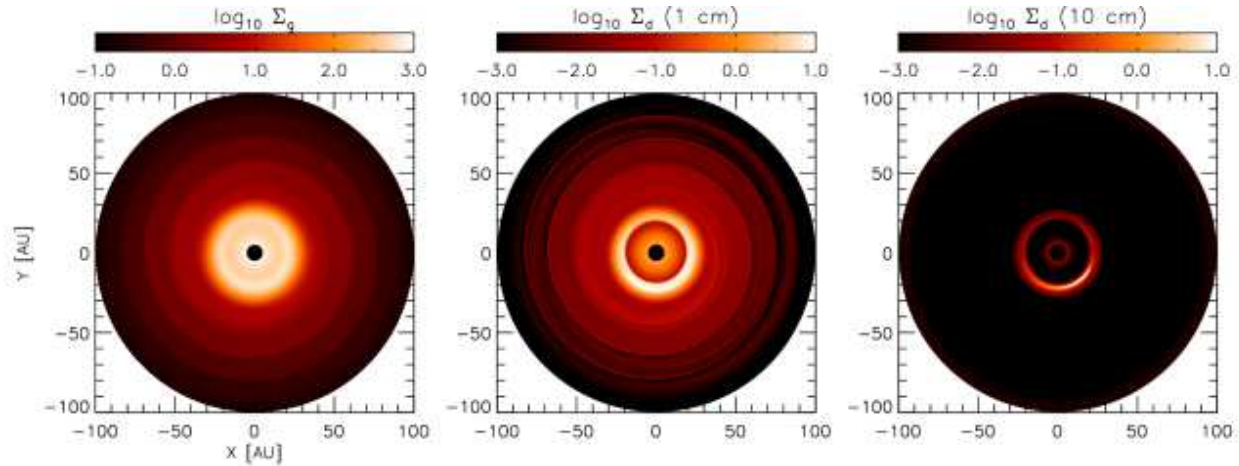}
\caption{Surface density distributions of (left) gas, (middle) 1~cm dust particles, and (right) 10~cm dust particles at the end of the simulation for the standard model. Densities are in cgs units and displayed in the logarithmic scale.}
\label{fig:figpolardens}
\end{figure*}

\section{DISCUSSION}

\subsection{Dust Trapping in Vortices and Observational Implications}
\label{sec:dust_trapping}

Our two-fluid approach shows that vortices formed by the RWI efficiently trap dust particles of appropriate size.
Figure \ref{fig:gtd} shows the time evolution of the minimum gas to dust ratio for the standard model.
Although the detailed evolution of the ratio for the small (1~cm) and large (10~cm) dust particles are  different, in overall, the ratios tend to decrease by more than a factor of few.
The small dust particles are well coupled to gas as we see in \S\ref{sec:dust} and therefore the decrease of the ratio is relatively smooth and only moderate.
The ratio has a minimum of $\sim20:1$.
For the large dust particles, the ratio slightly increases initially because they migrate inward rapidly (remember $T_s\sim 1$ at $R_c$), while gas surface density interior of $R_c$ increases due to the infall.
As vortices form, however, they efficiently trap the large dust particles and the gas to dust ratio drops to $\sim 2.7:1$ at $t\sim 120~T_{\rm orb}$.
The ratio then increases after $\sim 120~T_{\rm orb}$ but this is because the large dust at the outer disk are depleted due to rapid radial drift and there are no more supply from the region.

Figure \ref{fig:figpolardens} displays surface density distributions of gas, small dust, and large dust at the end of the standard run.
Clear dust concentration in the vortex is shown for both small and large dust particles.
Another notable feature is that the inner disk is depleted of dust because of the dust trapping by the vortex.
Again, this effect is more prominent for the larger dust particles.
As seen in the figure, existence of vortices can not only form azimuthal asymmetry of dust that can be directly observed via interferometric observations but also build radial structures that can be inferred by single-dish observations at multiple wavelengths. 
The deficit of dust at inner disk regions can affect the spectral energy distributions of the disk and such disks can be interpreted as transitional disks \citep{calvet05}.
The disks still can have enough gas at the inner disk (see Figure \ref{fig:figpolardens}) that can probably maintain high accretion rate observed in some transitional disks \citep[e.g.][]{espaillat10,espaillat11,andrews11,kim13}.

The significant dust concentration in vortices suggests that the vortices can provide favorable conditions for the planet and/or planetesimal formation.
In terms of vortex formation via the RWI, it would prefer higher infall rate and lower disk mass although a thorough parameter study is desired in the future regarding infall rate and disk mass.
This means that, with the example of HL Tau suggesting that planet formation can start early even during the infall phase, RWI might be one mechanism to accelerate planet formation.
On the other hand, the longevity of vortices has to be further tested since vortices may not survive forever \citep[e.g.][]{meheut12b}, although dust can still remain concentrated after gas vortices disappear \citep{birnstiel13}.

\subsection{Angular Momentum Transport}
\label{sec:transport}

\begin{figure*}
\centering
\epsscale{1}
\plotone{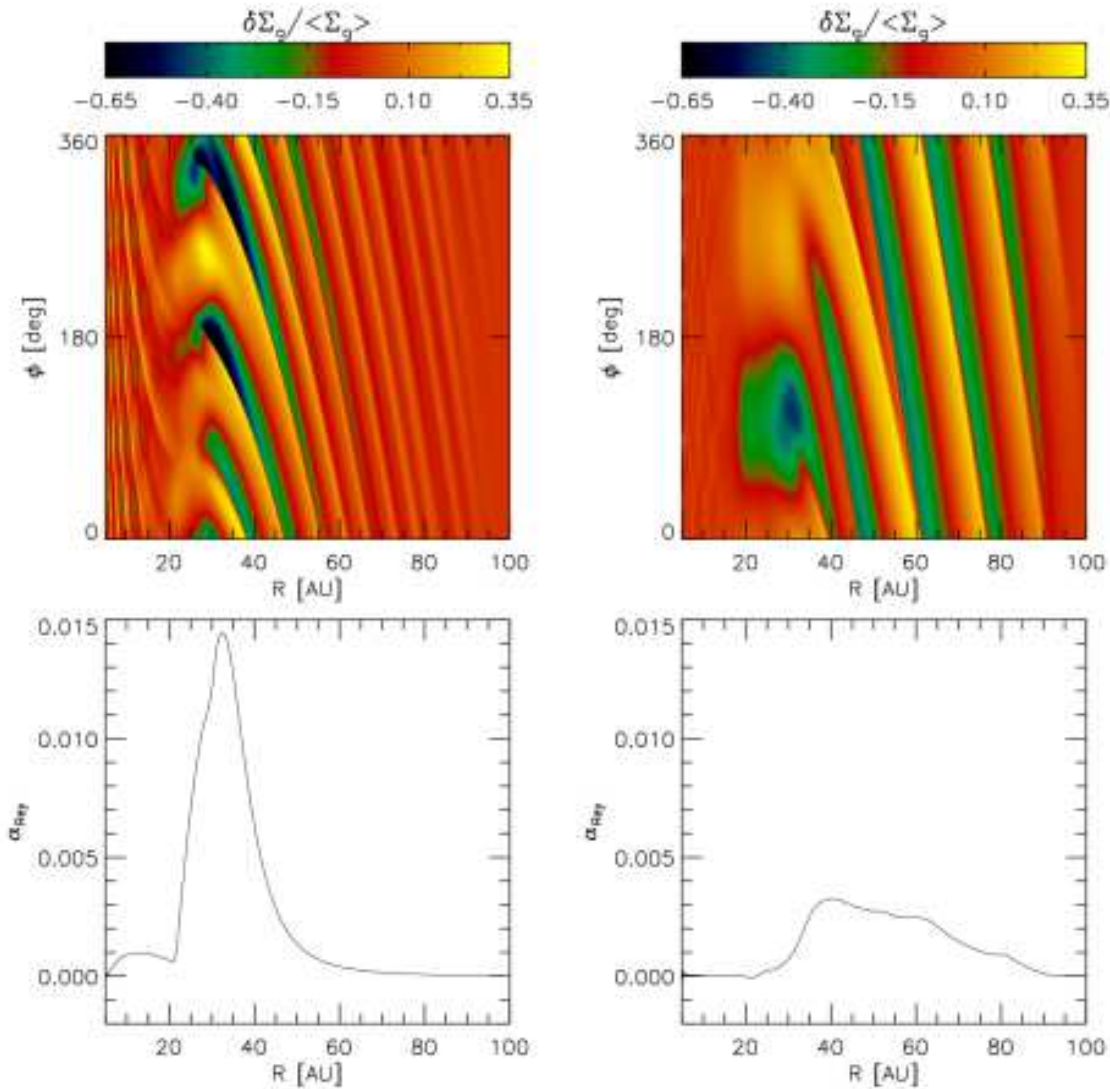}
\caption{(Top) Perturbed gas density $\delta \Sigma_g/ \langle \Sigma_g \rangle$ distributions on the $R-\phi$ coordinates. (Bottom) Azimuthally-averaged Reynolds stress as a function of radius. The snapshots are taken from the standard model at $t= 30~T_{\rm orb}$ (left) and $226~T_{\rm orb}$ (right).}
\label{fig:alpha}
\end{figure*}

\citet{li01} has shown that RWI-driven anticyclones and trailing spiral waves are responsible for outward angular momentum transport.
In Figure \ref{fig:alpha} we display spatial distribution of the perturbed gas density along with the radial distribution of azimuthally-averaged Reynolds stress.
At the time the instability saturates, the dominant $m=3$ mode is clearly seen at all radii: the trailing spiral waves propagate both interior and exterior of $R_c$.
The measured Reynolds stress has the maximum value of $\sim 0.015$ around $R_c$.
Inside of $R_c$ the Reynolds stress is measured to $\sim 10^{-3}$ while at $R>60$~AU the stress is in the range of $\alpha_{\rm Rey} \sim 10^{-4} - 10^{-6}$.
At later time, the vortices merge and the hydrodynamic turbulence around the vortex becomes less significant; $\alpha_{\rm Rey}\sim10^{-3}$.
Instead, the trailing one-armed spiral wave generate stronger turbulence very broadly in the disk that corresponds to $\alpha_{\rm Rey} \sim 10^{-4} - 3\times10^{-3}$.
The measured Reynolds stress driven by the RWI and the subsequent vortex formation is as strong as the ones induced by other instabilities in protoplanetary disks, such as gravitational instability \citep[e.g.][]{bae14} and magnetorotational instability \citep[e.g.][]{stone96}.

As a result of exchange of angular momentum between the vortex and the surrounding disk material, the vortex is subject to migration \citep{paardekooper10}.
More specifically, the vortex loses angular momentum and migrates inward if the trailing waves emitted by the vortex are stronger at the outer disk region than the inner disk region \citep{paardekooper10}.
We measure the migration rate starting from the time vortices merge together in a single, stable vortex, which is $t \sim 80~T_{\rm orb}$ in the standard model for example (see Figure \ref{fig:snap}).
However, we do not see the vortex migrating until the end of the simulation in our models.
It has to be noted though that this can be due to finite numerical resolution.
Since the radial grid size at $R_c$ is $\sim0.15$~AU, in the standard run we are able to observe migration only if the rate is $\gtrsim 6\times10^{-6}~{\rm AU~yr}^{-1}$.
We also do not observe vortex migrating in the higher resolution run with $1024 \times 2048$ grid cells, which reduces the migration rate to $\lesssim 3\times10^{-6}~{\rm AU~yr}^{-1}$.
The fact that the vortex migrates only very slowly, if it does, is important because the vortex might prevent centimeter to meter-sized objects from rapidly migrating inward in a few $\times 10^3 - 10^4$~years via the aerodynamic drag, retaining material for planet formation. 

Generally, it is known that vortices migrate toward high pressure regions \citep{paardekooper10,meheut12b}.
If a disk around a vortex has constant pressure, competing effects that determine vortex migration cancel each other and thus the vortex is likely to be locked in its location \citep{paardekooper10,lyra12,meheut12b}.
In our models, we find that the radial pressure gradient around merged vortices is shallow, in many cases close to zero, and this is presumably why we do not observe vortices migrating.
However, we caution that migration of vortices could depend on many other complications \citep[e.g.][]{richard13,faure15}, which are beyond the scope of two-dimensional adiabatic calculations.

\subsection{Caveats and Future Work}

It is likely that actual patterns of protostellar infall will be much more complex, possibly having filamentary infall pattern, as inferred from observations \citep[e.g.][]{tobin10,tobin11,tobin12,yen14} and suggested by numerical simulations \citep[e.g.][]{seifried15}.
These more complex patterns of mass and angular momentum addition might yield differing structures and other instabilities including RWI. 
Three-dimensional simulations are needed to take the next steps toward understanding the development of structure in protoplanetary disks.

It is possible that vortices generated during the protostellar infall phase dissipate during the subsequent disk evolution, especially under the influence of disk self-gravity as we show in \S\ref{sec:sg}.
In our example, however, the disk becomes gravitationally unstable quickly because we add infalling material at the same disk regions over time, whereas infall is likely to occur at larger radii as time passes, due to the addition of material with higher angular momentum \citep[e.g.][]{cassen81}.
Further studies of the longevity of vortices formed during the infall phase which incorporate a self-consistent and realistic evolutionary model are thus required.

Some other limitations of the current work include lack of the three-dimensionality of the RWI and vortex structure, proper thermodynamics which is especially important when self-gravity is considered, and more accurate dust physics such as dust growth and dust feedback.

\section{CONCLUSION}

We propose protostellar infall as a possible mechanism to trigger the RWI.
Our work demonstrates that the RWI enables early vortex formation in protoplanetary disks, during the infall phase.
This, along with the emerging observation evidences may suggest that planet/planetary core formation can start earlier than previously expected.

By implementing infall models, we carry out two-fluid, two-dimensional global hydrodynamic simulations.
Our results show that the RWI triggers at a radial minimum of the vortensity which is in agreement with previous works.
In our model the vortensity minimum develops near the density enhancement at the outer edge of the mass landing on the disk (centrifugal radius).
The key feature of triggering the RWI is the steep radial gradient of the azimuthal velocity close to $R^{-1}$ instead of $R^{-0.5}$ for Keplerian rotation, which is induced by the local increase in density at the centrifugal radius.
The instability initially grows in the linear regime where the growth is well described by the linear theory.
The vortensity minimum keeps growing as infall proceeds, and the instability eventually saturates, followed by subsequent nonlinear evolution.

We conduct a parameter study to investigate the RWI activity under a variety of disk conditions.
The major findings are as follows:
(1) the RWI triggers with disk viscosity of our interests ($\alpha \leq 10^{-2}$), with larger $\alpha$ weakening the RWI activity;
(2) the RWI tends to weaken if centrifugal radius increases over time;
(3) in case infalling material has faster inward radial velocity than the disk material on its landing, the vortensity minimum becomes sharper and narrower, resulting in more rapid growing of the RWI; 
(4) with a gentle mass addition where the radial infall pattern is matched on purpose to the initial disk surface density distribution, the RWI develops more slowly.

Vortex formation occurs when the instability enters the nonlinear phase.
Multiple vortices ($m \geq 3$) form initially but they merge to a single vortex in a few tens of local orbital time from their formation, in the absence of self-gravity.
Vortices generate trailing spiral waves that are responsible for outward angular momentum transport, with a Reynolds stress of $\lesssim 10^{-2}$.
However, no vortex migration was observed in our simulations.
Dust particles are well trapped in vortices in general, showing most prominent dust concentration for the particles with stopping times of the order of the orbital time ($T_s \sim 1$).
Dust trapping in vortices enhances the local dust to gas ratio significantly by a factor of $\sim40$ in our standard run.

With the evolutionary model and parameters we use in this study, the disk is gravitationally stable until the RWI saturates; thus the launching and growth of the RWI is not affected by self-gravity.
Due to continuous mass addition through infall and the presence of self-gravity, however, vortex merging tends to be impeded and the vortices eventually dissipate.

In this work, various disk conditions and infall patterns were tested sequentially to better isolate their effects on the RWI activity and the evolution of vortices.
The situation is much more complicated in actual protoplanetary disks.
In order for better understanding of the longer-term disk evolution, a more self-consistent and realistic long-term evolutionary model is desired in the future.
Also, long-term evolution and survival of vortices has to be tested under the consideration of disk turbulence and other instabilities that might be responsible for destroying vortices \citep[e.g.][]{kerswell02,lesur09}.

\acknowledgments

This research was supported in part through computational resources and services provided by Advanced Research Computing at the University of Michigan, Ann Arbor.
J.B. acknowledges Richard P. Nelson for carefully reading the manuscript and providing helpful comments.

\end{document}